\documentclass[twocolumn]{aastex631}

\begin{document}

\title{Elemental and Isotopic Yields from T Coronae Borealis: Predictions and Uncertainties}

\author[0009-0002-9117-7890]{Emma Wallace}
\affiliation{Department of Physics \& Astronomy, University of North Carolina at Chapel Hill, NC 27599-3255, USA}

\author[0000-0003-2381-0412]{Christian Iliadis}
\affiliation{Department of Physics \& Astronomy, University of North Carolina at Chapel Hill, NC 27599-3255, USA}
\affiliation{Triangle Universities Nuclear Laboratory (TUNL), Duke University, Durham, North Carolina 27708, USA}

\author[0000-0002-1359-6312]{Sumner Starrfield}
\affiliation{Earth and Space Exploration, Arizona State University, P.O. Box 871404, Tempe, AZ 85287-6004, USA}



\begin{abstract}
T~Coronae~Borealis (T~CrB) is a symbiotic recurrent nova system expected to undergo its next outburst within the next two years. Recent hydrodynamic simulations have predicted the nucleosynthetic yields for both carbon–oxygen (CO) and oxygen–neon (ONe) white-dwarf models, but without accounting for thermonuclear reaction-rate uncertainties. We perform detailed Monte Carlo post-processing nucleosynthesis calculations based on updated thermonuclear reaction rates and uncertainties from the 2025~evaluation. We quantify the resulting abundance uncertainties and identify the key nuclear reactions that dominate them. Our results show that both the CO and ONe nova models robustly produce characteristic CNO isotopes. More pronounced abundance differences emerge for elements with $A\ge20$. Sulfur is the most robust observational discriminator between the CO and ONe nova models, with a model-to-model difference of a factor of $\approx$30 and minimal sensitivity to reaction rate uncertainties. Neon, silicon, and phosphorus exhibit even larger abundance differences (factors of $\approx$150–250), providing strong diagnostic potential. While their predicted yields are subject to larger uncertainties, these remain smaller than the model-to-model differences, allowing these elements to serve as useful, though less precise, tracers of white-dwarf composition. Chlorine, argon, and potassium also differ between models, but the 1$\sigma$-abundance ranges for the CO and ONe models overlap, reducing their present usefulness as composition tracers. We find that only nine nuclear reactions dominate the abundance uncertainties of the most diagnostically important isotopes, and their influence is largely independent of the underlying white-dwarf composition. These results provide guidance for future experimental efforts and for interpreting ejecta compositions in the next eruption of T~CrB.
\end{abstract}



\section{Introduction} \label{sec:intro}
The symbiotic recurrent nova T Coronae Borealis (T~CrB), also known as the “Blaze Star,” is located less than $1$~kpc from Earth \citep{schaefer2022}. It belongs to the class of cataclysmic variables (CVs) and comprises a massive white dwarf, with a mass of approximately $1.35$~M$_\odot$ \citep{selvelli2019}, and a red giant companion in a close binary system with an orbital period of $\approx$ $228$~days \citep{kraft1958,Anupama1999}. The white dwarf gradually accretes hydrogen-rich material through Roche-lobe overflow from the red giant \citep{planqart25,munari2025tcrboverviewaccretion}. Once sufficient material has accumulated, a thermonuclear runaway is triggered on the white dwarf’s surface. This leads to the explosive ejection of both the accreted envelope and material dredged up from the white dwarf's outer layers. The elemental and isotopic composition of the ejecta provides a record of the nuclear processes that occurred during the explosion. Because thermonuclear reaction rates are highly sensitive to temperature \citep{Iliadis_2015}, the observed abundances offer crucial constraints for validating and refining hydrodynamic models of the outburst. After the eruption, the system settles back to its quiescent luminosity, and the cycle of mass transfer and eruption resumes.

Outbursts of T~CrB were observed in 1866 and 1946, during which the system reached a peak visual magnitude of $\approx$2. Additional historical evidence suggests earlier eruptions occurred in 1217 and 1787 \citep{schaefer2023}. Given the roughly 80-year recurrence interval as well as current observational indicators, T CrB is expected to erupt again within the next two years \citep{Schneider_2024}, so that observations with unprecedented detail will be possible, providing, for instance, precise determinations of the elemental and isotopic abundances of its ejected material.

Hydrodynamic simulations of the forthcoming outburst of T~CrB were presented by \citet{Starrfield_2025}. These simulations assumed matter accretion onto a $1.35$~M$_\odot$ white dwarf, with the goal of constraining the conditions necessary to trigger a thermonuclear explosion, following an inter-outburst interval of approximately 80 years. Both carbon–oxygen and oxygen–neon white dwarf compositions were considered to account for the potential diversity in ejecta abundances that may be observed in the next eruption. The simulations further suggested that the white dwarf in T~CrB is increasing in mass as a result of repeated thermonuclear runaways. Consequently, its long-term evolution may drive it toward the Chandrasekhar mass, possibly culminating in either a Type Ia supernova or an accretion-induced collapse, depending on its core composition. That study also predicted the nucleosynthetic yields of the nova ejecta, finding that all models substantially overproduced the intermediate-mass elements Si, P, and S, by several orders of magnitude relative to solar abundances. This is in addition to the well-known overproduction of odd-$A$ CNO isotopes such as $^{13}$C, $^{15}$N, and $^{17}$O \citep{Jose_1998}.

Another recent hydrodynamic study of the T~CrB outburst was presented by \citet{jose2025}. Those authors explored a wide range of hydrodynamic model parameters, including the white dwarf mass and luminosity, the metallicity of the accreted material, the mass-transfer rate, and the extent of mixing at the interface between the accreted envelope and the outer white dwarf layers. Their results revealed significant variations in the elemental abundances of numerous species, including Ne, Na, Mg, Al, Si, P, S, Ar, K, Ca, and Sc. These compositional differences were identified as potential diagnostics for constraining the underlying model parameters.

In the present work, we build upon the studies by \citet{Starrfield_2025} and \citet{jose2025} concerning the explosive nucleosynthesis expected in T~CrB, introducing several key advancements. First, while \citet{Starrfield_2025} and \citet{jose2025} employed thermonuclear reaction rates from the 2010 evaluation \citep{LONGLAND2010, ILIADIS2010b, ILIADIS2010c, iliadis2010d}, we adopt the updated and more accurate rates from the recent 2025 one \citep{iliadis25}. Second, neither \citet{Starrfield_2025} nor \citet{jose2025} provided uncertainties in the predicted abundances. In contrast, we quantify these uncertainties using the rate probability distributions from the 2025 evaluation. Third, we identify specific nuclear reactions whose rate uncertainties dominate the final abundance uncertainties of key elements and isotopes expected to be prominent in the ejecta of T~CrB. 

Assessing the role of reaction rate uncertainties is essential, as they contribute to the total uncertainty in predicted abundances,
adding to those arising from the hydrodynamic model parameters. Without quantifying the impact of nuclear physics uncertainties, it becomes difficult to disentangle their effects from those of the hydrodynamic inputs. Only by isolating and assessing the role of rate uncertainties can we determine whether observed elemental and isotopic abundances can meaningfully constrain the underlying physical conditions of the system.

The two nova models adopted in this study are described in Section~\ref{sec:explosion}. The reaction network and Monte Carlo simulations setup are detailed in Section~\ref{sec:simulations}. The resulting isotope abundances and the differences between the CO and ONe models are discussed in Sections~\ref{sec:finab} and~\ref{sec:diff}, respectively. The nuclear reactions most strongly impacting the final abundances are analyzed in Section~\ref{sec:reactions}. A summary of our key findings is provided in Section~\ref{sec:conclusion}.

\section{Explosion models} \label{sec:explosion}
To investigate the influence of reaction rates, we adopt the temperature-density evolutionary paths obtained from two hydrodynamical models for T~CrB presented by \citet{Starrfield_2025}. In these, the nova outburst is driven by accretion onto the white dwarf through Roche-lobe overflow from the companion red giant star into an accretion disk \citep{planqart25,munari2025tcrboverviewaccretion}. In terms of nuclear processes, near the peak temperature, most of the initially abundant nuclei in the CNO mass region are converted via proton captures into $\beta^+$-unstable nuclei, effectively halting any further increase in nuclear energy generation \citep{Starrfield1972}. All simulations exhibit a rapid rise in both temperature and nuclear luminosity to a maximum, followed by a more gradual decline. This rise in nuclear output coincides with the expansion of the convective region, which transports the $\beta^+$-decaying nuclei outward toward the surface, and simultaneously mixes unprocessed material downward into the active burning layers.

The models presented by \citet{Starrfield_2025} were computed using the one-dimensional (1D), fully implicit hydrodynamic code NOVA. From the 15 models explored in that study, consisting of five distinct initial compositions and three white dwarf radii, we selected two cases, assuming a white dwarf of CO or ONe composition, a mass of $1.35~M_\odot$, an initial luminosity of $L_{ini}$ $=$ $10^{-2}$~$L_\odot$, and a radius of $1522$~km \citep{Althaus2023}. These models adopt a mass accretion rate of $\dot{M}$ $=$ $1.1 \times 10^{-8}~M_\odot\mathrm{yr}^{-1}$, an accretion duration of $86.2$~yr, and a total accreted mass of $M_{\mathrm{acc}}$ $=$ $9.6 \times 10^{-7}~M_\odot$. Such an elevated accretion rate shortens the evolution time to the thermonuclear runaway (TNR) to approximately 80 years, roughly 100 times faster than the rates utilized in the classical CO and ONe nova studies of \citet{Starrfield_2020,Starrfield_2024}.

The first model adopted from \citet{Starrfield_2025} assumed a carbon–oxygen white dwarf (initially composed of 50\% $^{12}$C and 50\% $^{16}$O), whereas the second assumed an oxygen–neon composition. Both models utilized an initial composition for the thermonuclear burning consisting of 25\% white dwarf material mixed with 75\% solar matter \citep{Kelly_2013}. Values of the initial abundances for both models are listed in Table~2 of \citet{ward25}. The model properties are summarized in Table~\ref{tab:models}.

\begin{deluxetable}{lcc}
\tablecaption{Evolutionary characteristics of the models for T~CrB adopted in the present work.}
\label{tab:models} 
\tablewidth{\columnwidth}
\tablehead{
Property         & CO nova\tablenotemark{a}    & ONe nova\tablenotemark{a} 
}
\startdata
WD mass ($M_\odot$)                                 & 1.35                                 &     1.35    \\
WD composition                                      & $^{12}$C (50\%), $^{16}$O (50\%)     &     $^{16}$O, $^{20}$Ne,...\tablenotemark{d}   \\
Mixing (\%)\tablenotemark{b}                        & 25\% - 75\%                          &     25\% - 75\%    \\
$L_{ini}$~($L_\odot$)                               & 10$^{-2}$                                  & 10$^{-2}$     \\
$L_{peak}$~($10^4$ $L_\odot$)                       & 6.3                                  &     4.8     \\
$\tau_{acc}$~(yr)\tablenotemark{c}                  & 86.2                                 &     86.2    \\
$\dot{M}_{acc}$~($10^{-8}$ $M_\odot$ yr$^{-1}$)     & 1.1                                  &     1.1     \\
$M_{acc}$~($10^{-7}$ $M_\odot$)                     & 9.6                                  &     9.6     \\
$M_{ej}$~($10^{-8}$ $M_\odot$)                      & 5.6                                  &     4.6     \\
$M_{zone}$~($10^{-8}$ $M_\odot$)\tablenotemark{e}   & 5.4                                  &     5.4     \\
$T_{peak}$~(MK)                                     & 320                                  &     320     \\
\enddata
\tablenotetext{a}{From \citet{Starrfield_2025}.}
\tablenotetext{b}{The first and second percent values refer to white-dwarf and solar matter, respectively. Our initial abundances are listed in Table~2 of \citet{ward25}.}
\tablenotetext{c}{Time of accretion until TNR.}
\tablenotetext{d}{From mass point $1.17$~M$_\odot$ in \citet{ritossa1996}.}
\tablenotetext{e}{Mass of the zone extracted for the present one-zone simulations near the interface between the outer layers of the white dwarf and the accreted matter.}
\end{deluxetable}

Both models were computed assuming the accretion of solar-composition matter from the companion until a specific energy generation rate was achieved ($\approx$ $10^{11}$~erg~g$^{-1}$s$^{-1}$). At that point, the accretion was turned off, and the simulation continued by switching to a mixed composition (Table~2 of \citet{ward25}) through the peak of the TNR until the return to quiescence. This approach emulates the mixing process observed in multidimensional simulations, wherein white dwarf material is dredged upward into the accreted envelope once convective mixing is well-developed and extends close to the white dwarf's surface \citep{casanova2010, casanova2011}.

For each model, we focus on the hottest layer at the interface between the white dwarf's surface and the accreted envelope, as this is the region where most nuclear reactions are expected to occur \citep{jose2016stellar}. The temperature evolution of these zones is shown in the top panel of Figure~2 in \citet{Starrfield_2025}. Both models achieve the same peak temperature of $\approx 320$~MK (Table~\ref{tab:models}).

It is important to note that the final abundances from our post-processing simulations generally differ from those obtained in the full hydrodynamic models. First, our post-processing considers only a single zone and neglects the cooler outer layers. As will be shown, this leads to an overproduction of elements in the argon–potassium–calcium region, driven by the high peak temperatures reached in both models. In contrast, cooler outer zones would yield less material in this region and more in lighter elements such as phosphorus and sulfur. Second, convection is not included in the post-processing, meaning we neglect the mixing of material between zones. Third, the two simulations were performed using different nuclear reaction rate libraries, and some key rates have changed significantly between them. Despite these differences, we generally expect abundance ratios, particularly among closely related species, to show better agreement between the two approaches.

\section{Monte Carlo post-processing simulations}\label{sec:simulations}
We performed nucleosynthesis calculations using a reaction network consisting of 213 nuclides, ranging from protons, neutrons, and $^{4}$He up to $^{55}$Cr. These nuclides are interconnected by 2385 nuclear processes, including proton and $\alpha$-particle captures, $\beta$ decays, and various reactions involving lighter particles. Thermonuclear reaction rates were primarily obtained from STARLIB version 6.10 (December 2022). However, for 70 key reactions involving target nuclides in the mass range $A=2$–$40$, we adopted updated rates, uncertainties, and associated probability densities from the recent ``2025 Evaluation of Experimental Thermonuclear Reaction Rates" \citep{iliadis25}. These reactions cover most processes significant for nova nucleosynthesis. For a small number of reactions relevant to novae, experimental rates are not yet available, and STARLIB instead utilizes theoretical rates computed with the nuclear statistical model code TALYS \citep{Koning2023}. For these cases, an uncertainty factor of 10 was assumed.\footnote{Specifically, we adopted $f.u.=10$ for these reactions, and their rates were sampled using Equation~(\ref{eq:sample}).}

Stellar weak interaction rates, which depend on both temperature and density, were adopted from \citet{Fuller1982,Oda1994,NABI1999}. The stellar weak decay constants are tabulated at temperatures from $1$~MK to $10$~GK, and densities from $\rho Y_e$ $=$ $1$ to $10^{11}$~gcm$^{-3}$, where $\rho Y_e$ denotes the electron mole fraction. Short-lived nuclides, e.g., $^{13}$N ($T_{1/2}$ $=$ $10$~min), $^{14}$O ($T_{1/2}$ $=$ $71$~s), $^{15}$O ($T_{1/2}$ $=$ $122$~s), $^{17}$F ($T_{1/2}$ $=$ $64$~s), and $^{18}$F ($T_{1/2}$ $=$ $110$~min), present at the end of a network calculation were assumed to decay to their stable daughter nuclides. 

Each post-processing simulation commenced when the temperature in the hottest burning zone reached $100$~MK and continued until this zone was no longer connected by convection to the layers ultimately ejected during the outburst. For the models adopted in the present work, this corresponded to simulation times of $427$~s for the CO model and $9410$~s for the ONe one.

To quantify abundance uncertainties, we sampled simultaneously and randomly all reaction rates within the network. This sampling relies on the rate probability densities provided by the 2025 evaluation \citep{iliadis25}, or, where unavailable from that source, from STARLIB \citep{Sallaska2013}. We utilized a lognormal distribution for the thermonuclear reaction rate, $x$, of a given reaction, $j$, and temperature, $T$, represented by
\begin{equation}
\label{eq:ln}
f[x(T)_j] = \frac{1}{\sigma \sqrt{2\pi}}\frac{1}{x(T)_j} e^{-[\ln{x(T)_j} - \mu(T)_j]^2 / [2 \sigma(T)_j^2]}
\end{equation}
where the lognormal parameters $\mu$ and $\sigma$ determine the location and width, respectively. For a lognormal probability density, thermonuclear rate samples for reaction $j$ and network run $k$ are drawn using \citep{Longland2012}
\begin{equation}
\label{eq:sample}
x(T)_{kj} = x(T)_{med,j} [f.u.(T)]_j^{p(T)_{kj}}
\end{equation}
where $x_{med}$ and $f.u.$ are the median rate value and the rate factor uncertainty, respectively. Both of these quantities are listed in columns 2 and 3 of STARLIB, respectively. The variation exponent, $p_{kj}$, follows a normal distribution, meaning that it is characterized by a Gaussian distribution with a mean of zero and a standard deviation of one. It is important to note that the factor modifying the sampled reaction rate in comparison to its median value is $f.u.(T)^{p(T)}$, rather than $p(T)$. 

For a given network run, $k$, and nuclear reaction, $j$, we sample the variation exponent exactly once, meaning $p(T)_{kj}$ $=$ $p_{kj}$ remains constant across all temperatures \citep{Longland2012}. This procedure is repeated for a total of $n$ network runs to generate an ensemble of final abundance yields. The median abundance, $X_{\mathrm{med}}$, is defined as the 50th percentile of the resulting distribution, while the uncertainties are quantified by the 16th ($X_{\mathrm{low}}$) and 84th ($X_{\mathrm{high}}$) percentiles. The factor uncertainty for a given abundance is then calculated as $\Delta f$ $=$ $\sqrt{X_{\mathrm{high}} / X_{\mathrm{low}}}$, corresponding to a 68\% coverage probability.

In the following, we will discuss outcomes derived from conducting $5000$ Monte Carlo network simulations. Tests showed that this number is sufficiently large for statistical fluctuations to become smaller than the widths of the extracted abundance distributions. 

\section{Simulated abundances from the ejecta of the CO and ONe nova models}\label{sec:finab}
We begin by discussing abundance features common to both the CO and ONe models. Notable differences between the two will be addressed in the following section.

\begin{figure*}
\centering
\includegraphics[width=0.8\textwidth]{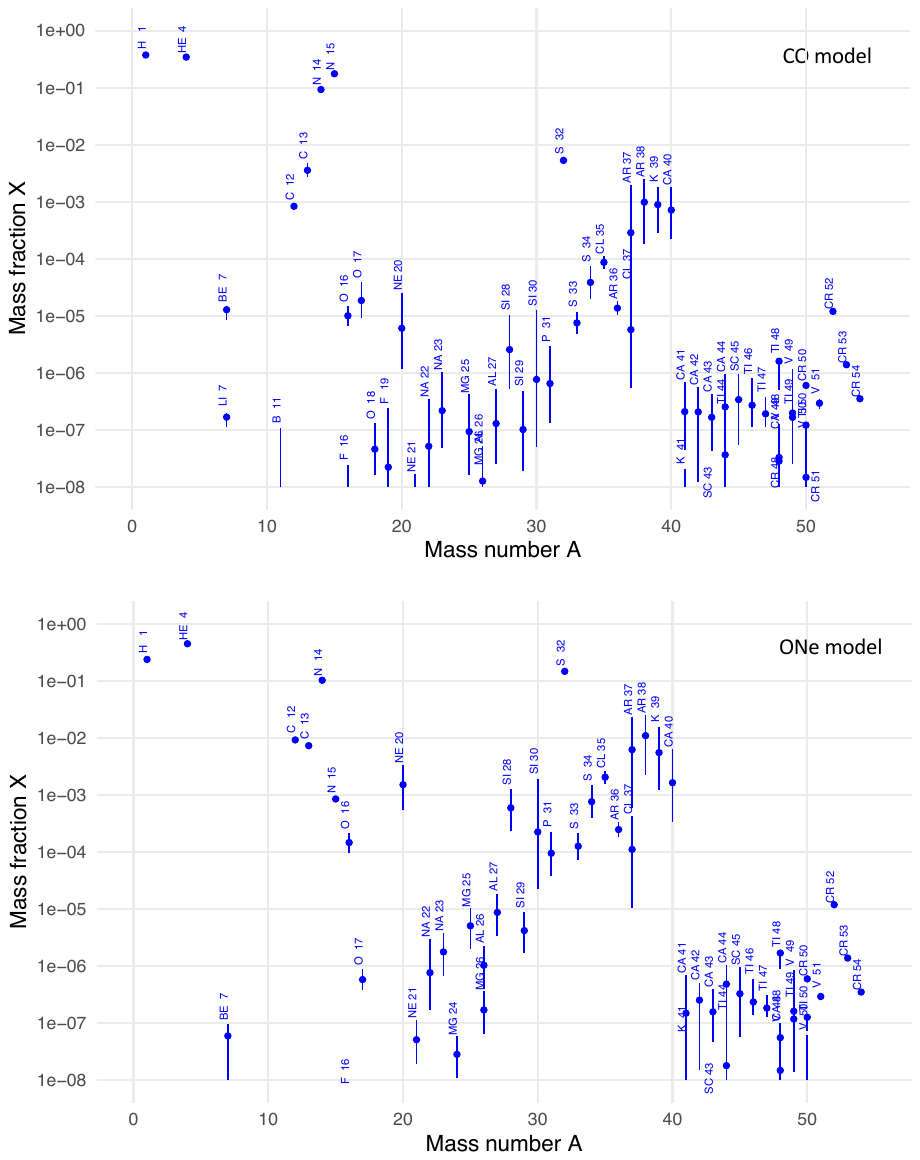}
\caption{Final mass fractions versus mass number produced by explosive burning in the hottest zone near the white dwarf surface. (Top) CO nova model; (Bottom) ONe nova model. The simulation predicts that the ejecta will be dominated, apart from hydrogen and helium, by $^{12}$C, $^{13}$C, $^{14}$N, and $^{15}$N in the CNO mass region, and by $^{32}$S, $^{37}$Ar, $^{38}$Ar, $^{39}$K, and $^{40}$Ca in the $A$ $\ge$ $20$ region.}
\label{fig:abund}
\end{figure*}

Simulated final mass fractions (columns 2 and 5 of Table~\ref{tab:models} for the CO and ONe nova model, respectively) as a function of mass number, resulting from explosive burning in the hottest zone near the white dwarf surface, are presented in Figure~\ref{fig:abund}. As expected, the CNO region of the ejecta is dominated by $^{12}$C, $^{13}$C, $^{14}$N, and $^{15}$N. The predicted nitrogen isotope mass fractions from our one-zone post-processing simulation agree quantitatively with those reported for the corresponding 1D hydrodynamic model by \citet{Starrfield_2025}, who used a multi-zone approach and averaged the results over the entire ejecta (see their Table~3, columns 3 and 5). The large $^{12}$C abundance reflects its substantial initial presence, owing to our assumption of significant $^{12}$C enrichment from the outer white dwarf layer in the accreted fuel. In contrast, the $^{13}$C, $^{14}$N, and $^{15}$N abundances originate from the $\beta$ decay of their respective progenitors, $^{13}$N, $^{14}$O, and $^{15}$O. These results are consistent with previous studies (e.g., \citet{Jose_1998}). Notably, the impact of thermonuclear reaction rate uncertainties on the predicted abundances is modest, typically less than 30\%. We also note that the $^{7}$Be abundance exceeds that of $^{7}$Li by several orders of magnitude  at the end of the network calculation (Section~\ref{sec:simulations}), consistent with the 1D hydrodynamic model results of \citet{Starrfield_2025}.

In the mass region $A$ $\ge$ $20$, the most abundant species are $^{32}$S, $^{37}$Ar, $^{38}$Ar, $^{39}$K, and $^{40}$Ca. As with the CNO mass region, our simulated mean mass fractions for $^{32}$S and $^{40}$Ca are qualitatively consistent with the 1D hydrodynamic results of \citet{Starrfield_2025} (see their Table~3, columns 3 and 5). A new result of the present work is the prediction of substantial abundances, $X_f$ $\ge$ $10^{-3}$ by mass, of $^{37}$Ar, $^{38}$Ar, and $^{39}$K, e.g., nuclides not included in the tabulation by \citet{Starrfield_2025}. Another new result from the present study is the estimation of associated uncertainties: while the abundance of $^{32}$S is predicted with relatively small uncertainty ($\lessapprox$7\%), the error bars for $^{38}$Ar, $^{39}$K, and $^{40}$Ca span about an order of magnitude, and those for $^{37}$Ar extend over two orders of magnitude. Notably, the predicted abundances of $^{37}$Ar and $^{38}$Ar exceed that of $^{36}$Ar, reported by \citet{Starrfield_2025} with values similar to ours, by several orders of magnitude.

\begin{figure*}
\centering
\includegraphics[width=0.8\textwidth]{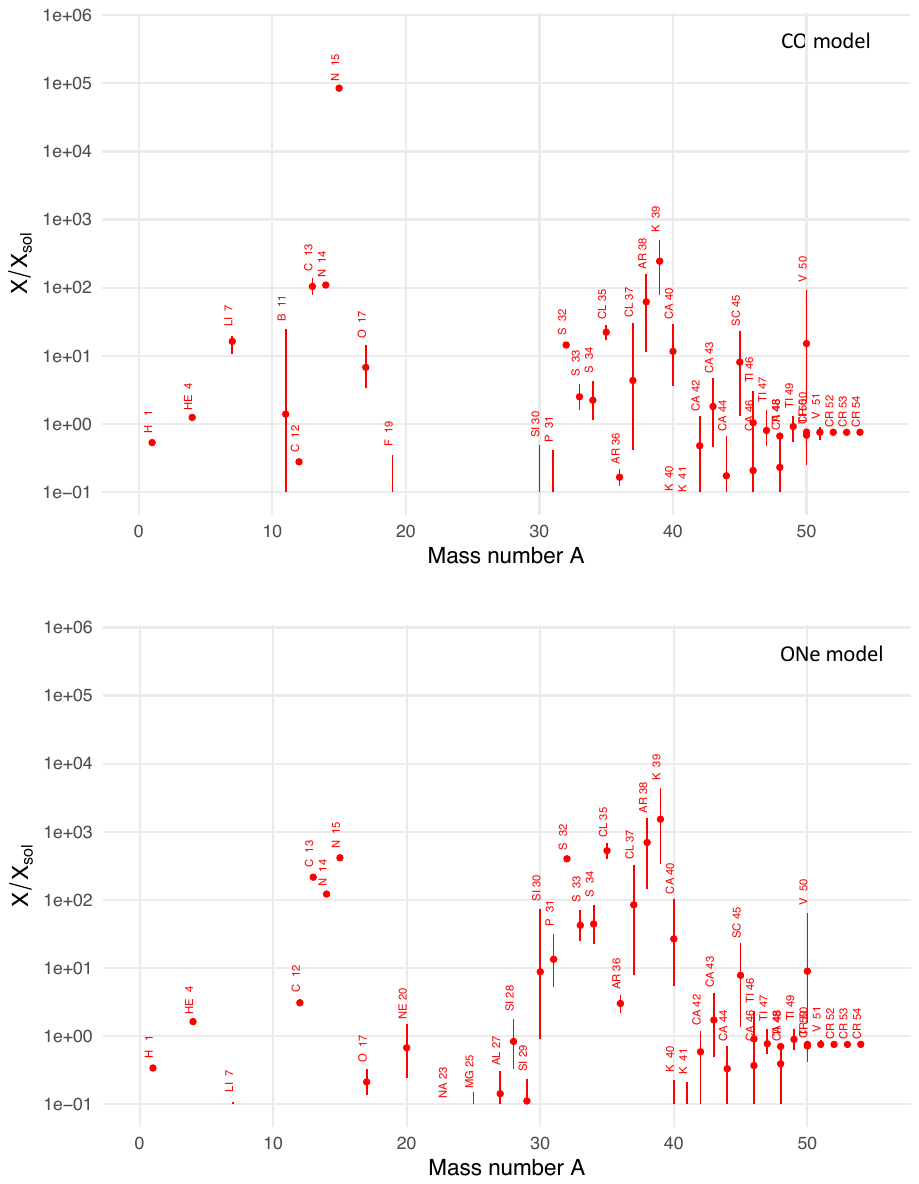}
\caption{Final mass fractions normalized to solar values \citep{Lodders_2021} versus mass number produced by explosive burning in the hottest zone near the white dwarf surface. (Top) CO nova model; (Bottom) ONe nova model. The simulation predicts that in the CNO mass region, the isotope $^{15}$N is overproduced by up to five orders of magnitude. In the $A$ $\geq$ $20$ region, the most overproduced isotopes are $^{38}$Ar and $^{39}$K, both exceeding their solar abundances by approximately two to three orders of magnitude.}
\label{fig:abundSolar}
\end{figure*}

Figure~\ref{fig:abundSolar} provides the results of the same simulation, but now displaying the final mass fractions normalized to solar values \citep{Lodders_2021} as a function of mass number. These abundances again reflect the products of explosive burning in the hottest zone near the white dwarf surface. In the CNO mass region, $^{15}$N is overproduced by several orders of magnitude, consistent with the findings of \citet{Starrfield_2025} (see their Figure~6), and its predicted abundance carries only a small uncertainty. In the $A$ $\geq$ $20$ region, the most overproduced isotopes are $^{38}$Ar and $^{39}$K, each exceeding their solar values by two (CO nova model) and three (ONe nova model) orders of magnitude. Again, the abundance uncertainties for these two species span roughly one order of magnitude. 

Other isotopes overproduced by at least an order of magnitude include $^{32}$S, $^{35}$Cl, $^{37}$Cl, $^{40}$Ca, $^{45}$Sc, and $^{50}$V in the CO nova model, and $^{30}$Si, $^{31}$P, $^{32}$S, $^{33}$S, $^{34}$S, $^{35}$Cl, $^{37}$Cl, $^{40}$Ca, $^{45}$Sc, and $^{50}$V in the ONe nova model. Among these, only $^{32}$S and $^{35}$Cl exhibit relatively small abundance uncertainties; the others display large uncertainties, some exceeding an order of magnitude. The abundance of $^{37}$Cl is primarily governed by the decay of $^{37}$Ar ($T_{1/2}$ $=$ $35$~d). Depending on the time of observation, a significant fraction of $^{37}$Ar may have already decayed to $^{37}$Cl.

\begin{deluxetable*}{cccccccc}
\tablecaption{Summary of simulated final abundances by mass in the hottest burning zone of the CO and ONe nova models. Results are only shown for those isotopes with mass fraction exceeding $\approx 10^{-3}$ or are overproduced compared to solar by at least a factor of $\approx 10$.\tablenotemark{a}}
\label{tab:results_1}
\tablewidth{0pt}
\tablehead{
  & \multicolumn{3}{c}{CO nova model}  & & \multicolumn{3}{c}{ONe nova model}  \\
\cline{2-4} \cline{6-8}
\colhead{Isotope} &  \colhead{$X_f$}  & \colhead{$X_f/X_\odot$}  &  \colhead{$\Delta f$}   & & \colhead{$X_f$}  & \colhead{$X_f/X_\odot$}  &  \colhead{$\Delta f$}   \\
     (1) & (2) & (3) & (4) & & (5) & (6) &  (7) 
}
\startdata
$^{12}$C                    &  8.33E-04   &   2.76E-01  &  1.08    &  &  9.25E-03    &    3.07E+00   &    1.12  \\
$^{13}$C                    &  3.55E-03   &   1.04E+02  &  1.32    &  &  7.32E-03    &    2.15E+02   &    1.09  \\
$^{14}$N                    &  9.27E-02   &   1.09E+02  &  1.01    &  &  1.03E-01    &    1.22E+02   &    1.02  \\
$^{15}$N                    &  1.75E-01   &   8.50E+04  &  1.002   &  &  8.52E-04    &    4.14E+02   &    1.06  \\
$^{20}$Ne                   &  6.03E-06   &   2.66E-03  &  4.7     &  &  1.52E-03    &    6.71E-01   &    2.5   \\
$^{28}$Si                   &  2.55E-06   &   3.55E-03  &  4.5     &  &  5.95E-04    &    8.31E-01   &    2.3   \\    
$^{30}$Si                   &  7.63E-07   &   2.96E-02  &  16      &  &  2.25E-04    &    8.75E+00   &    9.2   \\
$^{31}$P                    &  6.47E-07   &   9.10E-02  &  4.7     &  &  9.51E-05    &    1.34E+01   &    2.5   \\
$^{32}$S                    &  5.31E-03   &   1.44E+01  &  1.07    &  &  1.47E-01    &    4.00E+02   &    1.03  \\
$^{35}$Cl                   &  8.65E-05   &   2.22E+01  &  1.29    &  &  2.05E-03    &    5.27E+02   &    1.31  \\
$^{37}$Cl\tablenotemark{b}  &  5.70E-06   &   4.33E+00  &  8.5     &  &  1.11E-04    &    8.45E+01   &    6.3  \\
$^{37}$Ar                   &  2.86E-04   &   ...       &  8.5     &  &  6.24E-03    &    ...        &    6.3  \\
$^{38}$Ar                   &  9.80E-04   &   6.19E+01  &  3.7     &  &  1.10E-02    &    6.95E+02   &    3.3  \\
$^{39}$K                    &  8.89E-04   &   2.44E+02  &  2.5     &  &  5.54E-03    &    1.52E+03   &    3.6  \\
$^{40}$Ca                   &  7.13E-04   &   1.16E+01  &  2.8     &  &  1.64E-03    &    2.66E+01   &    4.4  \\
$^{45}$Sc                   &  3.38E-07   &   8.04E+00  &  4.2     &  &  3.27E-07    &    7.77E+00   &    4.1  \\
$^{50}$V                    &  1.46E-08   &   1.51E+01  &  6.5     &  &  8.68E-09    &    8.93E+00   &    6.5  \\
\enddata
\tablenotetext{a}{Columns: (1) Isotope label; (2), (5) Final mass fraction in hottest zone as a result of the explosion, at a time of one day after peak temperature; (3), (6) Final mass fraction relative to the solar value \citep{Lodders_2021} in hottest zone as a result of the explosion; (4), (7) Factor uncertainty, calculated as $\Delta f$ $=$ $\sqrt{X_{\mathrm{high}} / X_{\mathrm{low}}}$, corresponding to a 68\% coverage probability (see Section~\ref{sec:simulations}).}
\tablenotetext{b}{The abundance of $^{37}$Cl is due to the decay of $^{37}$Ar ($T_{1/2}$ $=$ $35$~d). Depending on the time of observation, $^{37}$Ar may have mostly decayed to $^{37}$Cl.}
\end{deluxetable*}

\section{Differences between the CO and ONe Nova models}\label{sec:diff}
The final elemental abundances of carbon, nitrogen, and oxygen in the hottest zone near the white dwarf surface are comparable, within an order of magnitude, for both the CO and ONe nova models. Once this material mixes with cooler outer layers during or after the thermonuclear runaway (TNR), these modest differences are expected to diminish. Therefore, our analysis emphasizes more substantial abundance differences between the models in the $A$ $\ge$ $20$ mass region. We again highlight isotopes with high final mass fractions ($X_f$ $\ge$ $10^{-3}$) or those overproduced relative to solar by at least an order of magnitude. As shown in Figure~\ref{fig:abund}, all isotopes of Ne, Si, P, S, Cl, Ar, and K exhibit enhanced abundances in the ONe nova model, typically by one or more orders of magnitude compared to the CO nova model.

The dominant neon isotope, $^{20}$Ne, exhibits markedly different behavior in the CO and ONe models. In the CO model, its final mass fraction is $X_f = 6.03 \times 10^{-6}$ (Table~\ref{tab:results_1}), several orders of magnitude below the solar value. In contrast, the ONe model begins with a high initial mass fraction of $\approx 8.2 \times 10^{-2}$ in the hottest zone, and decreases to $X_f \approx 1.5 \times 10^{-3}$ by the end of the thermonuclear runaway (TNR). This final abundance is close to the solar value ($X_\odot = 2.3 \times 10^{-3}$; see Figure~\ref{fig:abundSolar}) but still significantly lower than the elevated neon levels ($\approx 10^{-1}$) observed in some classical novae \citep[see Table 1 in][]{downen2013}. The resulting elemental neon abundance ratio between the CO and ONe models is 252, highlighting neon as a potentially valuable observational tracer for distinguishing between nova types. However, these simulated abundances carry notable uncertainties, by approximately a factor of 4.7 in the CO model and 2.5 in the ONe model.

Two isotopes mainly contribute to the elemental silicon abundance, $^{28}$Si and $^{30}$Si, which are produced in comparable amounts in both nova models. In the CO model, the final abundances of both isotopes remain below their solar values (see Table~\ref{tab:results_1}). In contrast, the ONe model yields a final $^{28}$Si abundance near the solar value, while $^{30}$Si is overproduced by a factor of 8.8. Overall, the total silicon production in the ONe model exceeds that of the CO one by a factor of $248$, with an uncertainty of about a factor of $6$. This difference suggests that measurements of elemental silicon could serve as a useful diagnostic for distinguishing between the two nova models.

A similar pattern is observed for $^{31}$P, which behaves like $^{30}$Si. Both isotopes are produced in similar quantities across the two nova models. In the ONe model, $^{31}$P reaches $X_f$ $\approx$ $10^{-4}$, which represents more than an order of magnitude overproduction relative to solar, and exceeds the CO model production by a factor of $\approx$150. The associated final-abundance uncertainties are a factor of 4.7 (CO model) and 2.5 (ONe model). 

Sulfur is the most abundant element in the $A$ $\ge$ $20$ mass range for both models. The dominant isotope, $^{32}$S, reaches final abundances of $X_f$ $=$ $5.3 \times 10^{-3}$ (CO model) and $X_f$ $=$ $1.5 \times 10^{-1}$ (ONe model), with small uncertainties ($\le 7$\%). These correspond to overproduction factors relative to solar of 14 and 400, respectively. The ONe-to-CO ratio of $\approx$30 suggests that sulfur may be a robust observational probe, as this distinction is not significantly affected by current reaction rate uncertainties.

For chlorine, $^{35}$Cl is overproduced by factors of 22 (CO model) and 530 (ONe model) relative to solar, indicating a 24-fold model-to-model difference. However, the situation is complicated by the contribution of $^{37}$Cl, which is fed by the decay of $^{37}$Ar ($T_{1/2}$ $=$ $35$~d). Since $^{37}$Ar is produced in even greater quantities than $^{35}$Cl after the TNR, the dominant chlorine isotope may vary depending on the observation time. Unfortunately, the uncertainties in the combined $^{37}$Cl and $^{37}$Ar abundances are substantial, factors of $8.5$ and $6.3$ for the CO and ONe models, respectively (see Table~\ref{tab:results_1}), resulting in overlapping 1$\sigma$ error bars. Thus, chlorine will only be a useful diagnostic if these uncertainties can be significantly reduced.

A similar issue arises for argon. The isotope $^{38}$Ar is overproduced by factors of $\approx 62$ (CO model) and $700$ (ONe model) compared to solar (see Table~\ref{tab:results_1}). Depending on the time of observation, $^{37}$Ar may be synthesized in comparable amounts to $^{38}$Ar, while $^{36}$Ar is produced at much lower levels. As with chlorine, the current uncertainties (Figure~\ref{fig:abund}) mean that the $^{37}$Ar and $^{38}$Ar abundances overlap within 1$\sigma$, limiting the usefulness of argon as a distinguishing probe unless uncertainties can be reduced.

The same general conclusion applies to potassium as well. Although $^{39}$K production is higher in the ONe model, the associated uncertainties, representing factors of $2.5$ (CO model) and $3.6$ (ONe model), again result in overlapping 1$\sigma$ ranges. Without reduced uncertainties, potassium cannot serve as a useful model discriminator.

Lastly, heavier isotopes such as $^{40}$Ca are produced with similar final abundances in both models and, therefore, are unlikely to serve as effective diagnostics for distinguishing between the CO and ONe nova models.

\section{Important nuclear reactions}\label{sec:reactions}
We now address the following question: Which current reaction rate uncertainties most strongly affect the predicted abundances of the isotopes discussed above? The influence of the rate uncertainty of a given reaction, $j$, on the outcome of nucleosynthesis can be determined by recording the $p_{kj}$ values for each sampled reaction network run, $k$ (see Section~\ref{sec:simulations}). A scatter plot displaying the final abundance of a specific nuclide against the sampled $p_{kj}$ values can then be analyzed to identify any correlations. 

In the past, Pearson’s (product-moment correlation coefficient) $r$, was used as a metric to quantify the correlation \citep[see, e.g.,][]{Coc_2014}. This is a measure of the linear correlation between two random variables. \citet{Iliadis2015} suggested using Spearman’s (rank-order correlation coeffcient) $r_s$, which quantifes how well the relationship between two variables is described by a monotonic function. However, both of these metrics are not without problems in the present context, where correlations are frequently neither linear nor monotonic.

To quantify which rates have the largest impact on a specific nuclidic abundance, we adopt instead the Mutual Information (MI) metric, which originates from information theory \citep{linfoot1957,coverthomas}. It quantifies the information one random variable (here the final mass fraction $X_f$ of a given nuclide for network run $k$) conveys about another (here the variation exponent $p_{kj}$), when both are sampled at the same time. For two random variables, $Y$ and $Z$, with values of $\{y_1, y_2, y_3,...\}$ and $\{z_1, z_2, z_3,...\}$, respectively, their Mutual Information is defined by
\begin{equation}
\label{eq:mi}
MI = \sum_y \sum_z P(y,z) \log \left[ \frac{P(y,z)}{P(y)P(z)} \right]
\end{equation}
where $P(y)$ and $P(z)$ are marginal distributions of $y$ and $z$, respectively, and  $P(y,z)$ is the joint probability density. A theorem from information theory states that the Mutual Information between two variables is zero if, and only if, the two random variables are statistically independent. 

\begin{figure*}
\centering
\includegraphics[width=1.0\textwidth]{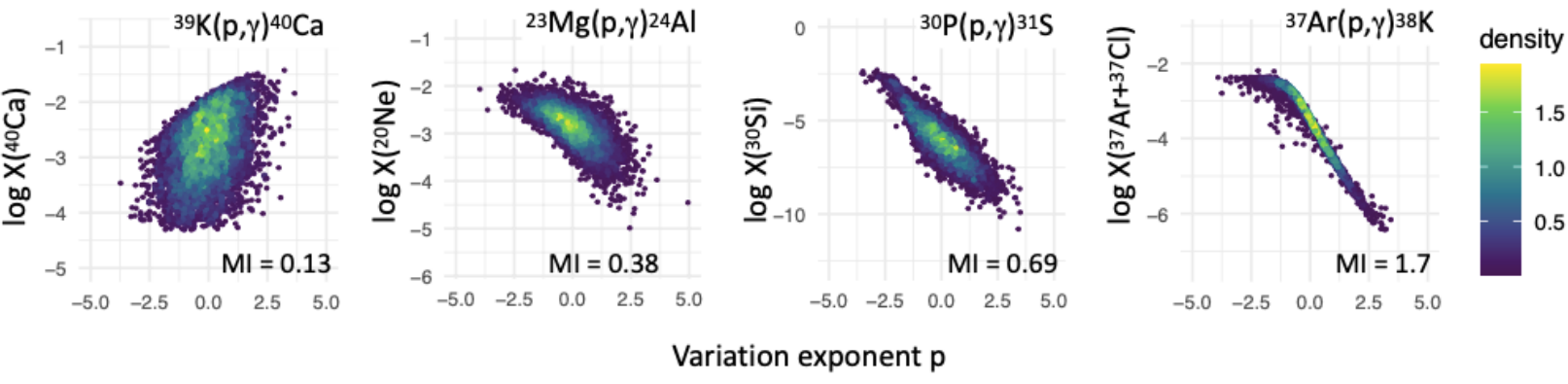}
\caption{Correlations of the simulated final abundance with the variation exponent, $p$, of a given reaction. Values for the Mutual Information (MI) are also given. The ordinate shows the logarithm of the mass fraction. From left to right: $^{40}$Ca abundance versus $^{39}$K(p,$\gamma$)$^{40}$Ca rate (ONe model); $^{20}$Ne abundance versus $^{23}$Mg(p,$\gamma$)$^{24}$Al rate (ONe model); $^{30}$Si abundance versus $^{30}$P(p,$\gamma$)$^{31}$S rate (CO model); $^{37}$Ar $+$ $^{37}$Cl abundance versus $^{37}$Ar(p,$\gamma$)$^{38}$K rate (CO model). 
}
\label{fig:corr}
\end{figure*}

Unlike conventional correlation metrics, the mutual information (MI) value depends not only on the strength of the relationship between variables, but also on the distribution of the data itself. As a result, MI provides a relative rather than an absolute measure of dependence. To facilitate interpretation, \citet{linfoot1957} proposed a normalization that transforms MI into the informational coefficient of correlation, defined as IC $\equiv$ $[1 - \exp{(-2 \times \textrm{MI})}]^{1/2}$. This coefficient ranges from 0 to 1, equaling zero for statistically independent variables and unity for fully correlated ones. In the present work, however, a normalization of MI is not necessary. Instead, we present in Figure~\ref{fig:corr} a set of correlation plots across a range of MI values, enabling the reader to visually gauge the corresponding strength of the dependence.

We focus on isotopes whose predicted abundance uncertainties exceed a factor of two: $^{20}$Ne, $^{28}$Si, $^{30}$Si, $^{31}$P, $^{37}$Cl, $^{37}$Ar, $^{38}$Ar, $^{39}$K, $^{40}$Ca, $^{45}$Sc, and $^{50}$V (see Table~\ref{tab:results_1}). In contrast, the abundance uncertainties for $^{12}$C, $^{13}$C, $^{14}$N, $^{15}$N, $^{32}$S, and $^{35}$Cl are smaller than 30\%—likely below the detection limits of future observational constraints for the ejecta of T~CrB. Furthermore, for each isotope considered, we include only those reactions with mutual information (MI) values exceeding 0.1. Below this threshold, the correlation between a reaction rate and the final abundance is generally too weak to be discerned visually (see Figure~\ref{fig:corr}).

Our key findings are summarized in Table~\ref{tab:reactions}. Remarkably, only nine distinct nuclear reactions are found to significantly influence the isotopes discussed above. Furthermore, the reactions that dominate the abundance uncertainties are nearly identical for both nova models considered, with comparable mutual information (MI) values in each case. This suggests that the sensitivity of the key isotopic abundances to reaction rate uncertainties is largely independent of the underlying white dwarf composition.

The final abundance of the isotope $^{20}$Ne is primarily affected by current uncertainties in the $^{23}$Mg(p,$\gamma$)$^{24}$Al rate, which yields mutual information (MI) values of 0.28 and 0.38 for the CO and ONe nova models, respectively; see the second panel in Figure~\ref{fig:corr}. This is because the reaction competes with the $\beta$-decay of $^{23}\mathrm{Mg}$ to $^{23}\mathrm{Na}$, thereby influencing how much material is channeled into $^{20}\mathrm{Ne}$ production via the $^{23}$Mg($\beta^+$)$^{23}\mathrm{Na}(\mathrm{p},\alpha)^{20}\mathrm{Ne}$ sequence versus being processed into heavier nuclei through $^{23}$Mg(p,$\gamma$)$^{24}$Al($\beta^+$)$^{24}$Mg(p,$\gamma$)$^{25}$Al. In the CO model, the $^{15}$O($\alpha$,$\gamma$)$^{19}$Ne reaction also contributes, albeit with a slightly lower MI value of 0.20, caused by a modest abundance flow from the CNO to the $A \ge 20$ mass region.

A similar situation arises for $^{28}$Si, where the same two reactions influence the final abundance, with comparable MI values. In contrast, the abundance of $^{30}$Si is strongly determined by a single reaction: $^{30}$P(p,$\gamma$)$^{31}$S, with MI values of $0.69$ (CO) and $0.98$(ONe); see the third panel in Figure~\ref{fig:corr}.

The final abundance of $^{31}$P is impacted by both the $^{23}$Mg(p,$\gamma$)$^{24}$Al and $^{30}$P(p,$\gamma$)$^{31}$S reactions, with MI values ranging from $0.16$ to $0.31$ across the two models. Additionally, in the CO model, the $^{15}$O($\alpha$,$\gamma$)$^{19}$Ne reaction contributes with an MI value of $0.19$.

As noted previously, the abundance of $^{37}$Cl is governed by the decay of $^{37}$Ar ($T_{1/2} = 35$~d), and the ratio of these isotopes depends on the time elapsed between peak temperature and observing the ejecta. The uncertainty in their combined abundance is dominated by a single reaction, $^{37}$Ar(p,$\gamma$)$^{38}$K, which exhibits high Mutual Information (MI) values of $1.7$ and $1.6$ for the CO and ONe nova models, respectively; see the fourth panel in Figure~\ref{fig:corr}.

The uncertainty in the $^{38}$Ar abundance is primarily driven by the $^{38}$K(p,$\gamma$)$^{39}$Ca reaction, with MI values of $0.76$ (CO) and $0.46$ (ONe). The $^{37}$Ar(p,$\gamma$)$^{38}$K reaction contributes as well, though to a lesser extent (MI $=$ $0.16$ and $0.33$, respectively).

The uncertainties in the $^{39}$K and $^{40}$Ca abundances are influenced by three reactions: $^{37}$Ar(p,$\gamma$)$^{38}$K, $^{38}$K(p,$\gamma$)$^{39}$Ca, and $^{39}$K(p,$\gamma$)$^{40}$Ca, with MI values ranging from $0.13$ to $0.28$; (see first panel in Figure~\ref{fig:corr}.

Finally, the uncertainties in the $^{45}$Sc and $^{50}$V abundances are dominated by the $^{45}$Sc(p,$\gamma$)$^{46}$Ti and $^{49}$Ti(p,$\gamma$)$^{50}$V reactions, respectively, for both nova models. For $^{45}$Sc, the $^{44}$Ca(p,$\gamma$)$^{45}$Sc reaction also contributes, but to a lesser extent.

\begin{deluxetable}{cccc}
\tablecaption{Summary of the most important reactions impacting the abundance uncertainties of the given isotopes in the CO and ONe nova models. Only isotopes whose abundance uncertainties are greater than a factor of two are included, and only reactions with MI values above 0.1 are presented. See text for details.\tablenotemark{a}}
\label{tab:reactions}
\tablewidth{0pt}
\tablehead{
&   &\multicolumn{2}{c}{MI value}    \\
\cline{3-4}
Isotope  &  Reaction & \colhead{CO model}  &    \colhead{ONe model}   \\
    (1) & (2)   & (3)       &  (4)
}
\startdata
$^{20}$Ne                               &  $^{23}$Mg(p,$\gamma$)$^{24}$Al        &   0.28    &   0.38      \\
                                        &  $^{15}$O($\alpha$,$\gamma$)$^{19}$Ne  &   0.20    &             \\
$^{28}$Si                               &  $^{23}$Mg(p,$\gamma$)$^{24}$Al        &   0.26    &   0.39      \\
                                        &  $^{15}$O($\alpha$,$\gamma$)$^{19}$Ne  &   0.20    &             \\
$^{30}$Si                               &  $^{30}$P(p,$\gamma$)$^{31}$S          &   0.69    &   0.98      \\
$^{31}$P                                &  $^{23}$Mg(p,$\gamma$)$^{24}$Al        &   0.24    &   0.31      \\
                                        &  $^{15}$O($\alpha$,$\gamma$)$^{19}$Ne  &   0.19    &             \\
                                        &  $^{30}$P(p,$\gamma$)$^{31}$S          &   0.16    &   0.22      \\
$^{37}$Cl + $^{37}$Ar\tablenotemark{b}  &  $^{37}$Ar(p,$\gamma$)$^{38}$K         &   1.7     &   1.6       \\
$^{38}$Ar                               &  $^{38}$K(p,$\gamma$)$^{39}$Ca         &   0.76    &   0.46      \\
                                        &  $^{37}$Ar(p,$\gamma$)$^{38}$K         &   0.16    &   0.33      \\
$^{39}$K                                &  $^{39}$K(p,$\gamma$)$^{40}$Ca         &   0.25    &             \\
                                        &  $^{37}$Ar(p,$\gamma$)$^{38}$K         &   0.22    &   0.26      \\
                                        &  $^{38}$K(p,$\gamma$)$^{39}$Ca         &   0.18    &   0.26      \\
$^{40}$Ca                               &  $^{38}$K(p,$\gamma$)$^{39}$Ca         &   0.28    &   0.27      \\
                                        &  $^{37}$Ar(p,$\gamma$)$^{38}$K         &   0.18    &   0.18      \\
                                        &  $^{39}$K(p,$\gamma$)$^{40}$Ca         &   0.17    &   0.13      \\
$^{45}$Sc                               &  $^{45}$Sc(p,$\gamma$)$^{46}$Ti        &   0.65    &   0.53      \\
                                        &  $^{44}$Ca(p,$\gamma$)$^{45}$Sc        &   0.41    &   0.63      \\
$^{50}$V                                &  $^{49}$Ti(p,$\gamma$)$^{50}$V         &   1.5     &   1.6       \\
\enddata
\tablenotetext{a}{Columns: (1) Isotope label; (2) Nuclear reaction(s) with largest impact on the abundance of the given isotope; (3) MI value for CO nova model; (4) MI value for ONe nova model (see Section~\ref{sec:reactions}).}
\tablenotetext{b}{The abundance of $^{37}$Cl is dominated by the decay of $^{37}$Ar (see comments in Table~\ref{tab:results_1}).}
\end{deluxetable}

\section{Conclusion}\label{sec:conclusion}
In this work, we have investigated the nucleosynthesis expected in the forthcoming nova outburst of T~CrB by building upon recent hydrodynamic simulations and incorporating the most up-to-date thermonuclear reaction rates and uncertainties from the 2025 evaluation. Using detailed Monte Carlo post-processing calculations, we quantified the impact of reaction rate uncertainties on predicted elemental and isotopic abundances, and identified the key reactions responsible for these uncertainties through Mutual Information analysis.

Our simulations reveal that both the CO and ONe nova models robustly produce characteristic CNO isotopes such as $^{13}$C, $^{15}$N, and $^{17}$O, in agreement with earlier studies. However, more substantial differences between the two models emerge in the $A$ $\ge$ $20$ mass region. 

Among all elements studied, sulfur, particularly $^{32}$S, remains the most promising observational discriminator between the CO and ONe nova models. It is produced at markedly different levels (a factor of $\approx$30) and exhibits very small uncertainties in both models ($\le$7\%), making it a robust probe of the underlying white dwarf composition. 

Neon, especially $^{20}$Ne, also shows a substantial model-to-model abundance difference (a factor of $\approx$250). Although the associated uncertainties are relatively large (factors of 4.7 and 2.5 in the CO and ONe models, respectively), they remain significantly smaller than the abundance ratio itself, preserving neon's utility as a diagnostic, albeit with reduced reliability compared to sulfur.

Silicon and phosphorus likewise show strong abundance differences between the models, with elemental production in the ONe model exceeding that in the CO model by factors of $\approx$250 and $\approx$150, respectively. However, these species are subject to larger uncertainties, up to a factor of $6$. The use of the silicon, phosphorus, and sulfur abundances in the ejecta of T~CrB as a diagnostic for distinguishing between nova models is also motivated by recent infrared observations of the recurrent novae U~Sco and V3890~Sgr, which revealed significant contributions of these elements in their ejecta \citep{Evans2022,Evans2023}.

In contrast, elements such as chlorine, argon, and potassium, while showing model-dependent abundance differences, suffer from large uncertainties that currently preclude their use as reliable discriminants. Reducing these abundance uncertainties through improved reaction rate measurements could expand the set of isotopes useful for constraining nova models observationally.

Finally, heavier species such as $^{40}$Ca and the classical CNO elements display little sensitivity to white dwarf composition, and thus are not useful discriminators.

In total, we identify only nine nuclear reactions that dominate the abundance uncertainties for the most diagnostically-relevant isotopes. These same reactions appear in both the CO and ONe models, suggesting that the sensitivity to nuclear physics uncertainties is largely independent of white dwarf composition. This finding reinforces the importance of experimental efforts to better constrain a relatively small number of reaction rates, most notably $^{30}$P(p,$\gamma$)$^{31}$S, $^{37}$Ar(p,$\gamma$)$^{38}$K, and $^{38}$K(p,$\gamma$)$^{39}$Ca, all of which exhibit Mutual Information (MI) values exceeding $0.5$.

Our results provide essential guidance for both additional nuclear astrophysics experiments and future observational campaigns. By identifying which isotopes are robust model discriminators and which nuclear reactions dominate their uncertainties, we take a critical step toward interpreting the expected ejecta composition from the next eruption of T~CrB, and, more broadly, toward using nova nucleosynthesis to constrain the physics of accreting white dwarfs.

\begin{acknowledgments}
We would like to thank D. P. K. Banerjee, A. Evans, T. Geballe, R. Janssens, U. Munari, and C. E. Woodward  for valuable feedback. CI acknowledges partial support from the DOE, Office of Science, Office of Nuclear Physics, under grants DE-FG02-97ER41041 (UNC) and DE-FG02-97ER41033 (TUNL), and his J. Ross Macdonald Distinguished Professorship. SS acknowledges partial support from a NASA Emerging Worlds grant to ASU (80NSSC22K0361) and his ASU Regents' Professorship.

\end{acknowledgments}

%







\bibliography{paper}{}
\bibliographystyle{aasjournal}



\end{document}